\documentclass[showpacs,preprintnumbers,prd,nofootinbib,floats,amssymb,floatfix]{revtex4}
\usepackage{amsmath}
\usepackage{amsthm}
\usepackage{graphicx}
\usepackage{amsxtra}
\usepackage{hyperref}
\usepackage{amssymb}
\usepackage{amstext}
\begin{document}
\title{ New canonical  analysis for higher order  topologically massive gravity }
\author{Alberto Escalante}  \email{aescalan@ifuap.buap.mx}
\author{Jorge Hern\'andez Aguilar}  \email{ 219470418@alumnos.fcfm.buap.mx}
 \affiliation{Instituto de F\'isica ``Luis Rivera Terrazas'', Benem\'erita Universidad Aut\'onoma de Puebla.
 \\
 Apartado Postal J-48 72570, Puebla Pue., M\'exico, }
 \affiliation{Facultad de Ciencias F\'isico Matem\'aticas, Benem\'erita Universidad Aut\'onoma de Puebla.}
\begin{abstract}
A detailed Gitman-Lyakhovich-Tyutin   analysis for higher-order topologically massive gravity is performed. The  full structure of the constraints, the counting of physical degrees of freedom, and the Dirac algebra among the constraints are reported. Moreover, our analysis presents a new structure in the constraints and we compare our results with those reported in the literature where a standard Ostrogradski framework was developed.

\end{abstract}
 \date{\today}
\preprint{}
\maketitle
\section{Introduction}
Nowadays the study of higher-order theories is an interesting subject in theoretical physics. In fact, since the works developed by Ostrogradski concerning the  Hamiltonian formulation of such systems \cite{1}  up to now where  recent works regarding systems just like generalizations of electrodynamics \cite{2, 3, 4, 5}, generalized meson-field theory \cite{5a, 5b}, string theory \cite{6a, 6}, dark energy physics \cite{7, 8} have  been analyzed. Moreover, from the gravitational point of view, we find interesting models of higher derivative gravity where  are useful for testing the investigation of quantum gravitational effects. Those models are formed with the Einstein-Hilbert action and the addition of quadratic products of the curvature tensor; the main attractive feature of those models is the renormalizability \cite{9, 10}.
Within the context of three dimensions, we find an interesting and natural higher-order model, the so-called topologically massive gravity [TMG] \cite{11a}. This theory is  the coupling of Einstein-Hilbert action plus a Chern-Simons term, at a linear level it describes the propagation of a single massive state of helicity $\pm$ 2 on a Minkowski background, the so-called massive graviton. The theory is a good laboratory for testing classical and quantum ideas of gravity because it is the unique local dynamical and unitary gravity model which is naive power counting renormalizable \cite{11c}. In this respect, the study of  higher-order field theories is carried out by using the so-called Ostrogradski-Dirac framework \cite{11b}. In fact, the Ostrogradski-Dirac framework is based on the extension of the phase space, where  the choice of the fields and their temporal derivatives are the canonical variables. In other words, a generalization of the canonical momenta  is introduced, then the identification of the constraints is performed as in Dirac's method is done \cite{11}. However, in some cases it is claimed that the usual Ostrogradski-Dirac framework does not have the control for identifying the constraints in consistent form, then the constraints are fixed by hand \cite{12a}. In this respect, a detailed Hamiltonian formulation should provide a complete description of the system, and the complete set of either constraints or gauge invariance transformations must be correctly identified. Hence, it is mandatory to perform a  correct analysis  of the constraints in any   canonical formulation. In this respect, with all constraints at hand and classified into first class and second class, the Dirac brackets, useful to quantize a gauge system can be constructed, then the second-class constraints and non-physical degrees of freedom are removed. In the case of higher-order gravity theories, the separation of the constraints into first-and second-class is a delicate issue, therefore, alternative approaches can be required. In this respect, there exists another, alternative, a canonical formulation of higher-order systems the so-called Gitman-Lyakhovich-Tyutin [GLT] formalism \cite{14, 15}. The GLT formalism is a generalization of Ostrogradski's framework that is based on the introduction of extra fields reducing a problem with higher derivatives to one with first-order time derivative, then by using either a correct gauge fixing and the introduction of the Dirac brackets, the second class constraints and non-physical degrees of freedom can be removed. \\
In this manner, with all commented above, the purpose of the present work is to report a detailed GLT study of TMG theory. Our analysis will follow a different procedure to that presented in \cite{15a}. We  will develop a detailed GLT framework as an alternative study beyond Ostrogradski-Dirac framework. In fact, the correct canonical analysis of a given classical theory is the first step towards its canonical quantization and so it is worthwhile to perform it. The  complete structure of the constraints without fixing them by hand  are reported. In addition, we will compare our results with those reported in the literature.\\
The paper is organized as follows. In Section II the GLT method for higher-order Chern-Simons theory is developed. We report the complete set of the constraints and the Dirac brackets are constructed; the Dirac algebra between the constraints and the first-class extended Hamiltonian is reported. In Section III the TMG theory is analyzed. We perform a detailed GLT formalism and the new structure of the constraints is reported. In addition, the Dirac brackets and the algebra of the constraints are calculated.
    
\section{GLT analysis of higher order Chern-Simons term}
We start with the standard linear  form of the Chern-Simons Lagrangian density  given by \cite{15a}  
\begin{equation}
    L_{CS}=\frac{1}{2}\epsilon^{\lambda \mu \nu} (\partial_\sigma h^{\rho}{}_{\lambda} \partial_\rho \partial_\mu h^{\sigma}{}_{\nu}-\partial_\sigma h^{\rho}{}_{\lambda} \partial^\sigma \partial_\mu h_{\rho \nu}),
\label{eq:lcs}    
\end{equation}
here, spacetime indices are represented by  the greek alphabet $\alpha, \beta = 0, 1, 2$ and space indices  by the latin   $i, j, k =1, 2$, $h_{\mu \nu}$ is the perturbation of the metric around the flat spacetime geometry and the following signature  $\eta_{\mu \nu}= (-1, 1, 1)$ is used. By performing the $2+1$ decomposition, we can write the action as
\begin{equation}
\begin{aligned}
    L_{CS}= {} & - \epsilon^{ij}(\partial^k h_{0j}\Ddot{h}_{ki} + \partial_j h^{k}{}_{0}\Ddot{h}_{ki} - \frac{1}{2} \dot{h}^{k}{}_{j}\Ddot{h}_{ki} + \partial_j \partial^k h_{00} \dot{h}_{ki}+                \partial^k \partial_i h_{0j} \dot{h}_{k0} \\ 
               & + \frac{1}{2} \nabla^2 h_{0j} \dot{h}_{0i} - \frac{1}{2} \nabla^2 h^{k}{}_{j} \dot{h}_{ki} + \frac{1}{2} \partial_k \partial_l h^{l}{}_{j} \dot{h}^{k}{}_{i} + \nabla^2 h_{00} \partial_i h_{0j} \\
               & - \nabla^2 h^{k}{}_{0} \partial_i h_{kj} - \partial^l \partial_i h^{k}_{j} \partial_k h_{l0}).
\end{aligned}
\label{eq:cssep}
\end{equation}
We can observe that the Lagrangian is a higher-order derivative theory and the standard way for performing the canonical analysis is by using  the Ostrogradski-Dirac framework, however, we have commented above the interest to develop an alternative GLT analysis. It is worth commenting that our results are new and are not reported in the literature.  For our aims, we need to rewrite the Lagrangian by introducing the following variables 
\begin{equation}
    G_{\mu \nu} = \dot h_{\mu \nu}
 \quad \quad 
    v_{\mu \nu} = \dot G_{\mu \nu}=\Ddot{h}_{\mu \nu},
\label{eq:var}
\end{equation}
hence, inserting   (\ref{eq:var}) into  (\ref{eq:cssep}) and introducing the momenta $(\pi ^{\mu \nu},  P^{\mu \nu})$  canonically conjugate  to $(h_{\mu \nu}, G_{\mu \nu})$, we redefine the action as \cite{14, 15}

\begin{equation}
    S= \int{\left[ L_{CS}^{v} + \pi ^{\mu \nu}(\dot h_{\mu \nu} - G_{\mu \nu}) + P^{\mu \nu}(\dot G_{\mu \nu} - v_{\mu \nu})\right]{d}^{3}x},
\label{eq:actioncs}    
\end{equation}
where 
\begin{equation}
\begin{aligned}
    L_{CS}^{v}= {} & - \epsilon^{ij}(\partial^k h_{0j} v_{ki} + \partial_j h^{k}{}_{0} v_{ki} - \frac{1}{2} G^{k}{}_{j} v_{ki} + \partial_j \partial^k h_{00} G_{ki} + \partial^k \partial_i h_{0j} G_{k0} \\ 
               & + \frac{1}{2} \nabla^2 h_{0j} G_{0i} - \frac{1}{2} \nabla^2 h^{k}{}_{j} G_{ki} + \frac{1}{2} \partial^k \partial_l h^{l}{}_{j} G_{ki} + \nabla^2 h_{00} \partial_i h_{0j} \\
               & - \nabla^2 h^{k}{}_{0} \partial_i h_{kj} - \partial^l \partial_i h^{k}{}_{j} \partial_k h_{l0}), 
\end{aligned}
\label{eq:lagcst}    
\end{equation}
we can observe that there is an advantage of the action  (\ref{eq:actioncs}) because it contains only first-order time derivatives of the fields. It is worth mentioning that the introduction of the momenta allows us to identify more easily the constraints in comparison with Ostrogradski's formalism \cite{14, 15}. In fact, in GLT framework it is not necessary the introduction of a generalized canonical momenta for the fields with  higher-order time derivative terms just as in Ostrogradski is done. Furthermore, there  is not any  effect in the counting of degrees of freedom because  (\ref{eq:cssep}) and (\ref{eq:actioncs}) are equivalent (see the appendix A). In this respect, the fundamental Poisson brackets are given by
\begin{eqnarray}
\nonumber 
    \left\lbrace h_{\alpha \beta},\pi^{\mu \nu}\right\rbrace &=&\frac{1}{2}\left(\delta^{\mu }{}_{\alpha} \delta^{\nu }{}_{\beta} + \delta^{\mu }{}_{\beta} \delta^{\nu }{}_{\alpha}\right), \\ 
    \left\lbrace G_{\alpha \beta},P^{\mu \nu}\right\rbrace&=&\frac{1}{2}\left(\delta^{\mu }{}_{\alpha} \delta^{\nu }{}_{\beta} + \delta^{\mu }{}_{\beta} \delta^{\nu }{}_{\alpha}\right).
\label{eq:eq20}
\end{eqnarray}
In this manner,  the canonical Hamiltonian is defined as usual  
\begin{equation}
\begin{aligned}
    H_{CS}={} & \pi^{\mu \nu } G_{\mu \nu } + P^{\mu \nu } v_{\mu \nu } - L^v_{CS}  \\
= {} &  \pi^{00} G_{00} + P^{00} v_{00} + (2\pi^{0k} + \frac{1}{2}\epsilon^{kj}\nabla^2 h_{0j} + \epsilon^{ij}\partial^k \partial_i h_{0j})G_{0k}\\
                & + [\pi^{ki} + \epsilon^{ij}(\partial_j \partial^k h_{00} - \frac{1}{2} \nabla^2 h^{k}{}_{j} + \frac{1}{2} \partial^k \partial_l h^{l}{}_{j})]G_{ki} + 2P^{0i}v_{0i}\\
                & + [P^{ki} + \epsilon^{ij}(\partial^k h_{0j} + \partial_j h^{k}{}_{0} - \frac{1}{2} G^{k}{}_{j} )]v_{ki} \\
                & + \epsilon^{ij} ( \nabla^2 h_{00} \partial_i h_{0j} - \nabla^2 h^{k}{}_{0} \partial_i h_{kj} - \partial^l \partial_i h^{k}{}_{j} \partial_k h_{l0} ),
\end{aligned}
\label{eq:HCS}
\end{equation}
where (\ref{eq:var}) and (\ref{eq:lagcst}) have been used. Thus,  from the  Lagrangian (\ref{eq:lagcst}) we   identify the following primary constraints \cite{14, 15}
\begin{equation}
    Q^{\mu \nu} : \frac {\partial L_{CS}^{v}}{\partial v_{\mu \nu }} - P^{\mu \nu } \approx 0,
\end{equation}
say 
\begin{equation}
\begin{aligned}
    Q^{ki} {} & : \frac {\partial L_{CS}^{v}}{\partial v_{ki}} - P^{ki} \approx 0\\ & = - \left \{P^{ki} +\frac{1}{2} \left [ \epsilon^{ij}(\partial^k h_{0j} + \partial_j h^{k}{}_{0} - \frac{1}{2} G^{k}{}_{j} ) + \epsilon^{kj}(\partial^i h_{0j} + \partial_j h^{i}{}_{0} - \frac{1}{2} G^{i}{}_{j} ) \right ] \right \} \approx 0,
\end{aligned}
\label{eq:cntrns}
\end{equation}
\begin{equation}
    Q^{00}: \frac {\partial L_{CS}^{v}}{\partial v_{00}} - P^{00}=0 \Rightarrow P^{00}\approx 0,
\label{eq:Res1}
\end{equation}
\begin{equation}
    Q^{0i}: \frac {\partial L_{CS}^{v}}{\partial v_{0i}} - P^{0i}=0 \Rightarrow P^{0i} \approx 0.
\label{eq:Res2}
\end{equation}
The Poisson algebra between the primary constraints is given by 
\begin{equation}
    \left\lbrace Q^{00},Q^{00} \right\rbrace = \left\lbrace Q^{00},Q^{0i}\right\rbrace = \left\lbrace Q^{0i},Q^{0i} \right\rbrace \approx 0,
\label{PoissonQQ}
\end{equation}
\begin{equation}
     \left\lbrace Q^{ki},Q^{lm} \right\rbrace=\frac{1}{4} \left (  \epsilon^{mi}\eta^{lk} + \epsilon^{li}\eta^{mk} + \epsilon^{lk}\eta^{mi} + \epsilon^{mk}\eta^{li} \right ), 
\end{equation}
and the primary Hamiltonian is defined by 
\begin{equation}
    H^{1}_{CS}:=H_{CS}+ \Delta_{\mu \nu}Q^{\mu \nu},
\label{eq:H1}
\end{equation}
where  \(\Delta_{\mu \nu}\) are idetified  as Lagrange multipliers. In this manner, from consistency of primary constraints 
\begin{equation}
    \dot Q^{\alpha \beta}=\left\lbrace Q^{\alpha \beta},H_{CS}^{1} \right\rbrace=\left\lbrace Q^{\alpha \beta},H_{CS} \right\rbrace + \Delta_{\mu \nu}\left\lbrace Q^{\alpha \beta},Q^{\mu \nu} \right\rbrace \approx 0, 
\label{eq:cnsevol}
\end{equation}
we identify the following secondary constraints 
\begin{equation}
   S^{0}: \left\lbrace Q^{00},H_{CS} \right\rbrace= \pi^{00}\approx 0 ,
\end{equation}
\begin{equation}
   S^{i}: \left\lbrace Q^{0i},H_{CS} \right\rbrace=  \left ( \pi^{0i} + \frac{1}{4}\epsilon^{ij}\nabla^2 h_{0j} + \frac{1}{2}\epsilon^{lj} \partial^i \partial_l h_{0j} \right ) \approx 0,
\end{equation}
and relations between the  Lagrange multipliers 
\begin{equation}
\begin{split}
   \left\lbrace Q^{ki},H_{CS} \right\rbrace  & = \frac{1}{2} \bigg [ \epsilon^{ij} \left (\partial_j \partial^k h_{00} - \frac{1}{2} \nabla^2 h^{k}{}_{j} + \frac{1}{2} \partial^k \partial_r h^{r}{}_{j} \right ) + \epsilon^{kj} \left ( \partial_j \partial^i h_{00} - \frac{1}{2} \nabla^2 h^{i}{}_{j} + \frac{1}{2} \partial^i \partial_r h^{r}{}_{j} \right ) \bigg ] \\ &\quad  + \pi^{ik} - \frac{1}{2} \left ( \epsilon^{lk}v^{i}_{l} + \epsilon^{li}v^{k}_{l} \right) - \frac{1}{2} \bigg [ \epsilon^{ij} \left ( \partial^{k} G_{0j}+\partial_{j} G^{k}_{0} \right ) + \epsilon^{kj} \left ( \partial^{i} G_{0j}+\partial_{j} G^{i}_{0} \right ) \bigg ] \\   &\quad - \frac{1}{2} \left (  \epsilon^{mk} \Delta^{i}_{m} + \epsilon^{mi} \Delta^{k}_{m}  \right ) \approx 0.
\end{split}
\label{eq:evolqik}
\end{equation}
The Poisson algebra between the secondary constraints is given by 
\begin{equation}
    \left\lbrace S^{0},S^{0} \right\rbrace =
     \left\lbrace S^{\alpha},Q^{00} \right\rbrace = \left\lbrace S^{\alpha},Q^{0i}\right\rbrace = \left\lbrace S^{0},Q^{ik}\right\rbrace = 0,
\label{PoissonSS}
\end{equation}
thus, consistency of the secondary constraints allows us to  identify the following tertiary  constraint 
 \begin{equation}
\dot S^{0} \rightarrow    \tilde{S}^0 : \left [ \epsilon^{ij} \partial_{j} \partial^{k} G_{ki} + \epsilon^{ij} \nabla^{2} ( \partial_{i} h_{0j})  \right ]\approx0,
\end{equation}
and  $\dot S^{i} $  provides  relations between the Lagrange multipliers  
\begin{equation}
\begin{split}
\dot S^{i} \rightarrow    \tilde{S}^i :&  \frac{\epsilon^{ik}}{2} \left (\nabla^{2} G_{0k} + \partial^{l} \partial_{k} G_{0l} - \partial^{l}v_{lk} - \partial_{k}\nabla^{2}h_{00} \right ) \\ &\quad + \frac{\epsilon^{kj}}{2} \left ( \partial_{j} v^{i}{}_{k} + \nabla^{2}\partial_{k}h^{i}{}_{j} - \partial_{l} \partial^{i}\partial_{k}h^{l}{}_{j} + \partial^{i}\partial_{k}G_{0j} \right ) +\frac{1}{2} \left( \epsilon^{ki}\partial^{l}\Delta_{lk} + \epsilon^{kj}\partial_{j}\Delta^{i}{}_{k}  \right) \approx0.
\end{split}
\label{eq:evolsi}
\end{equation}
From temporal evolution  of the tertiary constraint $\tilde{S}^0$,    we do not obtain more constraints because the Lagrange multipliers are mixed  
\begin{equation}
    \dot {\tilde{S}}^{0} = \epsilon^{ij} \left ( \partial_{j}\partial^{k}v_{ki} + \nabla^{2}\partial_{i}G_{0j} + \partial_{j}\partial^{k}\Delta_{ki} \right)\approx 0. 
\end{equation}
On the other hand,  this is not the end,  because there are more constraints. In fact, we can observe that  the trace of  (\ref{eq:evolqik})  eliminate the Lagrange multipliers and therefore, we identify   other   constraint 
\begin{equation}
V=\pi^i{_{i}} +\frac{\epsilon^{ij}}{2} \partial_i \partial_lh^l{_{j}} \approx 0.
\label{60c}
\end{equation}
Moreover, from (\ref{eq:evolqik})  and  (\ref{eq:evolsi}) we eliminate the Lagrange multipliers and other  constraints are obtained 
\begin{equation}
\begin{split}
    \tilde{V}^{i} & = \partial_{k}\pi^{ik} + \frac{\epsilon^{kl}}{4}\nabla^2\partial_kh^i{_{l}}- \frac{\epsilon^{kj}}{4}\partial_k \partial^i\partial_lh^l{_{j}} \approx 0. 
\end{split}
\label{59c}
\end{equation}
Finally, we observe that the trace of  (\ref{eq:cntrns})  implies other constraint and from consistency  do not emerge further constraints, say 
\begin{eqnarray}
        & \eta_{ik}Q^{ik}& = P^{i}{}_{i}, \nonumber \\
        &\dot P^{i}{}_{i}& = \left\lbrace P^{i}{}_{i},H_{CS} \right\rbrace = \tilde{V} \approx 0. 
\end{eqnarray}
Therefore the complete set of constraints is given by 
\begin{eqnarray}
&Q^{00}&:P^{00} \approx0, \nonumber \\
&Q^{0i}&:P^{0i} \approx0, \nonumber \\
&Q^{ik}&:P^{ki} +\frac{\epsilon^{ij}}{2}  \partial^k h_{0j} + \frac{ \epsilon^{ij}}{2}\partial_j h^{k}{}_{0} - \frac{\epsilon^{ij}}{4} G^{k}{}_{j}  + \frac{\epsilon^{kj}}{2} \partial^i h_{0j} + \frac{\epsilon^{kj}}{2}\partial_j h^{i}{}_{0} - \frac{\epsilon^{kj}}{4} G^{i}{}_{j} \approx0,  \nonumber \\
&S^{0}& :   \pi^{00}  \approx 0, \nonumber \\
&S^{i}&:    \pi^{0i} + \frac{\epsilon^{ij}}{4}\nabla^2 h_{0j} + \frac{\epsilon^{lj}}{2}\partial^i \partial_l h_{0j}  \approx 0, \nonumber\\
 &\tilde{S}^0& : \epsilon^{ij} \partial_{j} \partial^{k} G_{ki} + \epsilon^{ij} \nabla^{2}  \partial_{i} h_{0j} \approx 0, \nonumber \\
  &\tilde{V}^{i} & :\partial_{k}\pi^{ik} + \frac{\epsilon^{kl}}{4}\nabla^2\partial_kh^i{_{l}}- \frac{\epsilon^{kj}}{4}\partial_k \partial^i\partial_lh^l{_{j}} \approx 0, \nonumber \\
 & \tilde{V}&:\pi^i{_{i}} +\frac{\epsilon^{ij}}{2} \partial_i \partial_lh^l{_{j}} \approx 0, \nonumber \\
 & U&: P^{i}{}_{i} \approx 0.
 \label{fullcons}
\end{eqnarray}
Now, the  not trivial algebra between all constraints is expressed by 
\begin{equation}
W = \bordermatrix{
                   & Q^{00}    & Q^{0l}   & Q^{lm}   & S^{ 0}   & S^{l}   & \tilde{S}^{0}   & \tilde{V}^{l}   & \tilde{V}   &U   \cr
    Q^{00}         & 0         & 0        & 0        & 0        & 0        & 0               & 0               & 0        & 0        \cr
    Q^{0i}         & 0         & 0        & 0        & 0        & 0        & 0               & 0               & 0        & 0        \cr
    Q^{ik}         & 0         & 0        & \left\lbrace Q^{ik},Q^{lm} \right\rbrace        & 0        & \left\lbrace Q^{ik},S^{l} \right\rbrace & \left\lbrace Q^{ik},\tilde{S}^{0} \right\rbrace                & 0               & 0         & 0       \cr
    S^{ 0}         & 0         & 0        & 0        & 0        & 0        & 0               & 0               & 0         & 0       \cr
    S^{ i}         & 0         & 0        & \left\lbrace S^{i},Q^{lm} \right\rbrace        & 0        & \left\lbrace S^{i},S^{l} \right\rbrace        & \left\lbrace S^{i},\tilde{S}^{0} \right\rbrace               & 0               & 0         & 0       \cr
    \tilde{S}^{0}  & 0         & 0        & \left\lbrace \tilde{S}^{0},Q^{lm} \right\rbrace        & 0        & \left\lbrace \tilde{S}^{0},{S}^{l} \right\rbrace         & 0               & 0               & 0         & 0       \cr
    \tilde{V}^{i}  & 0         & 0        & 0        & 0        & 0        & 0               & 0            & 0           & 0     \cr
    \tilde{V}      & 0         & 0        & 0        & 0        & 0        & 0               & 0            & 0           & 0     \cr
    U              & 0         & 0        & 0        & 0        & 0        & 0               & 0            & 0           & 0     \cr 
            }.
    \label{eq:WCS}
\end{equation}
where 
\begin{eqnarray}
& \left\lbrace Q^{ik},Q^{lm} \right\rbrace & = \frac{1}{4} \left (  \epsilon^{mi}\eta^{lk} + \epsilon^{li}\eta^{mk} + \epsilon^{lk}\eta^{mi} + \epsilon^{mk}\eta^{li} \right )\delta^{2}(x-y), \nonumber \\
& \left\lbrace Q^{ik},S^{l} \right\rbrace & =  \frac{1}{4} \left ( \epsilon^{il}\partial^{k} + \epsilon^{kl}\partial^{i} + \epsilon^{ij}\eta^{kl}\partial_{j} + \epsilon^{kj}\eta^{il}\partial_{j} \right )\delta^{2}(x-y), \nonumber \\
& \left\lbrace Q^{ik},\tilde{S}^{0} \right\rbrace & = \frac{1}{2} \left ( \epsilon^{ji}\partial_{j}\partial^{k} + \epsilon^{jk}\partial_{j}\partial^{i} \right ) \delta^{2}(x-y), \nonumber \\
& \left\lbrace S^{i},S^{l} \right\rbrace & = \frac{1}{4} \left ( \epsilon^{il}\nabla^{2} + \epsilon^{ij}\partial^{l}\partial_{j} + \epsilon^{jl}\partial^{i}\partial_{j} \right )\delta^{2}(x-y), \nonumber \\
& \left\lbrace S^{i},\tilde{S}^{0} \right\rbrace & = \frac{1}{2}\epsilon^{li}\nabla^{2}\partial_{l}\delta^{2}(x-y),
\end{eqnarray}
from the matrix (\ref{eq:WCS})  we can classify  the constraints into first class and second class. In fact,  from  the rank of this matrix  we identify the set of second class constraints,  and  the  null vectors allows us to  identify the first class constraints  \cite{18}, thus,  from the null vectors  we identify the following first class  constraints
    \begin{eqnarray}
&Q^{00}&:P^{00} \approx0, \nonumber \\
&Q^{0i}&:P^{0i} \approx0, \nonumber \\
&S^{0}& :   \pi^{00}  \approx 0, \nonumber \\
&S^{i}&:   \partial_{k}P^{ki} - \pi^{0i} + \frac{\epsilon^{ij}}{2}\partial_{k}\partial_{j}h^{k}{}_{0} +  \frac{\epsilon^{ij}}{4}\nabla^{2}h_{0j} - \frac{\epsilon^{ij}}{4}\partial^{k}G_{kj} - \frac{\epsilon^{kj}}{4}\partial_{k}G^{i}_{j} \approx0 \nonumber\\
&\tilde{V}^{i} & :\partial_{k}\pi^{ik} + \frac{\epsilon^{kl}}{4}\nabla^2\partial_kh^i{_{l}}- \frac{\epsilon^{kj}}{4}\partial_k \partial^i\partial_lh^l{_{j}} \approx 0, \nonumber \\
 & \tilde{V}&:\pi^i{_{i}} +\frac{\epsilon^{ij}}{2} \partial_i \partial_lh^l{_{j}} \approx 0, \nonumber \\
 & U&: P^{i}{}_{i} \approx 0,\nonumber \\
 &\tilde{S}^0& \partial_{i}\partial_{k}P^{ik} + \frac{\epsilon^{ij}}{2} \partial_{i} \partial^{k} G_{kj},
\end{eqnarray}
and the rank implies  the following two  second class constraints 
    \begin{eqnarray}
& Q^{11}& = P^{11} + \partial^{1}h_{02} + \partial_{2}h^{1}{}_{0} -\frac{1}{2}G^{1}{}_{2} \approx 0, \nonumber\\
& Q^{12}& = P^{12} + \partial^{2}h_{02} - \partial^{1}h_{01} + \frac{1}{4}G^{1}{}_{1} - \frac{1}{4}G^{2}{}_{2}  \approx 0, 
    \end{eqnarray}
we can observe that the constraints are identified by means of the rank and the null vectors;  it is not necessary to fixing  them by hand such as in Ostrogradski formalism is done \cite{15a}.  For instance, one null vector is given by $w=(0, 0,  \partial_i \omega, 0, \eta_{i}{^{j} }\omega, 0, 0, 0)$, where $\omega$ is an arbitrary function. From the contraction of the vector $w$  with the constraints (\ref{fullcons}), the first class constraint   $S^i$  is obtained. \\
With the correct constraints,  we are able to calculate the number of physical degrees of freedom. In fact, there are 24 canonical variables, 11 first class constraints and two second class constraints, this allows us to conclude that
\begin{equation}
DOF = \frac{1}{2}\left[ \left ( 24 \right ) -2\left ( 11\right )- \left ( 2\right )  \right] = 0,
\end{equation}
as expected.\\
To complete the analysis, it is well-known that for two functions on the phase space, namely $A$ and $B$, the Dirac bracket between these variables is defined by
\begin{equation}
    \left\lbrace A,B \right\rbrace_{D} =  \left\lbrace A,B \right\rbrace - \int dudv  \left\lbrace A,\chi^{\alpha} (u) \right\rbrace C^{-1}_{\alpha \beta} \left\lbrace \chi^{\beta} (v),B \right\rbrace,  
    \label{eq:DIRACB}
\end{equation}
where we have noted to  $\chi^{\alpha}$ and  $\chi^{\beta}$ as the second class constraints,   and  $C^{-1}_{\alpha \beta}$  is the inverse matrix of $C^{\alpha\beta}$ whose entries are the Poisson brackets between the second class constraints. Thus, for the theory under study  $C^{\alpha\beta}$  takes the form
\begin{equation}
    C^{\alpha \beta} = \begin{pmatrix}
        \left\lbrace Q^{11},Q^{11} \right\rbrace &\left\lbrace Q^{11},Q^{12} \right\rbrace  \\
        \left\lbrace Q^{12},Q^{11} \right\rbrace & \left\lbrace Q^{12},Q^{12} \right\rbrace
    \end{pmatrix} = \begin{pmatrix} 
        0 & -\frac{1}{2} \\
        \frac{1}{2} & 0
    \end{pmatrix}\delta^{2}(x-y),
\end{equation}
and its inverse is given by 
\begin{equation}
    C^{-1}_{\alpha \beta} = \begin{pmatrix} 
        0  & 2 \\
        -2 & 0
    \end{pmatrix}\delta^{2}(x-y).
\end{equation}
Therefore, the non trivial  Dirac's  brackets between the canonical variables are given  by 
\begin{eqnarray}
&\left\lbrace h_{00},\pi^{00} \right\rbrace_{D}& = \delta^{2}(x-y), \nonumber \\
&\left\lbrace h_{0i},\pi^{0l} \right\rbrace_{D}& = \frac{1}{2}\delta_{l}{}^{i}\delta^{2}(x-y), \nonumber \\
&\left\lbrace h_{ij},\pi^{lm} \right\rbrace_{D}& = \frac{1}{2} \left( \delta_{i}{}^{l}\delta_{j}{}^{m} + \delta_{i}{}^{m}\delta_{j}{}^{l}\right) \delta^{2}(x-y), \nonumber \\
&\left\lbrace \pi^{0i},\pi^{0l} \right\rbrace_{D}& = -\frac{1}{2} \epsilon^{il}\nabla^{2} \delta^{2}(x-y), \nonumber \\
&\left\lbrace \pi^{0i},G_{lm} \right\rbrace_{D}& = \frac{1}{2} \left( \delta_{l}{}^{1}\delta_{m}{}^{2} + \delta_{l}{}^{2}\delta_{m}{}^{1} \right) \left( \epsilon^{1i}\partial^{1} +  \eta^{1i}\partial^{2} \right) \delta^{2}(x-y) \nonumber \\
&& -\frac{1}{2}\delta_{l}{}^{1}\delta_{m}{}^{1} \left( \epsilon^{1i}\partial^{2} +  \epsilon^{2i}\partial^{1} + \eta^{2i}\partial^{2} - \eta^{1i}\partial^{1} \right) \delta^{2}(x-y), \nonumber \\
&\left\lbrace \pi^{0i},P^{lm} \right\rbrace_{D}& = -\frac{1}{8}\left(  \epsilon^{ln}\eta^{im} + \epsilon^{mn}\eta^{il} + \epsilon^{mi}\eta^{ln} + \epsilon^{li}\eta^{mn}  \right)\partial_{n} \delta^{2}(x-y), \nonumber \\
&\left\lbrace G_{00},P^{00} \right\rbrace_{D}& = \delta^{2}(x-y), \nonumber \\
&\left\lbrace G_{0i},P^{0l} \right\rbrace_{D}& = \frac{1}{2}\delta_{i}{}^{l}\delta^{2}(x-y), \nonumber \\
&\left\lbrace G_{ij},G_{lm} \right\rbrace_{D}& = \delta_{i}{}^{1}\delta_{j}{}^{1}\left( \delta_{l}{}^{1}\delta_{m}{}^{2} + \delta_{l}{}^{2}\delta_{m}{}^{1} \right)\delta^{2}(x-y) - \delta_{l}{}^{1}\delta_{m}{}^{1}\left( \delta_{i}{}^{1}\delta_{j}{}^{2} + \delta_{i}{}^{2}\delta_{j}{}^{1} \right)\delta^{2}(x-y), \nonumber \\
&\left\lbrace G_{ij},P^{lm} \right\rbrace_{D}& = \frac{1}{4}\delta_{i}{}^{1}\delta_{j}{}^{1}\left(  \epsilon^{1m}\eta^{2l} + \epsilon^{1l}\eta^{2m} + \epsilon^{2m}\eta^{1l} + \epsilon^{2l}\eta^{1m}\right)\delta^{2}(x-y) \nonumber \\
&& -\frac{1}{4}\left( \delta_{i}{}^{1}\delta_{j}{}^{2} + \delta_{i}{}^{2}\delta_{j}{}^{1}\right)\left( \epsilon^{1m}\eta^{1l} + \epsilon^{1l}\eta^{1m} \right)\delta^{2}(x-y) \nonumber \\
&& +\frac{1}{2} \left( \delta_{i}{}^{l}\delta_{j}{}^{m} + \delta_{i}{}^{m}\delta_{j}{}^{l}\right) \delta^{2}(x-y), \nonumber \\
&\left\lbrace P^{ij},P^{lm} \right\rbrace_{D}& = \frac{1}{16}\left( \epsilon^{il}\eta^{mj} + \epsilon^{im}\eta^{jl} + \epsilon^{jl}\eta^{mi} + \epsilon^{jm}\eta^{il} \right)\delta^{2}(x-y).
\label{dbra}
\end{eqnarray}
Finally, the correct classification of the constraints allows us to construct the extended Hamiltonian. In fact, it is well known that the extended Hamiltonian must to be a first class function. Moreover,  the extended Hamiltonian and the Dirac brackets are the cornerstones for performing the identification of either observables or the study of quantization. The extended Hamiltonian is given by
\begin{equation}
H_{E}=H_{CS}+u_{\alpha} \chi^{\alpha},
\label{HTCS}
\end{equation}
where $u_{\alpha} $ are the Lagrange multipliers that can be obtained from 
\begin{equation}
    u_{\alpha}= C^{-1}_{\beta \alpha}\lbrace \chi^{\beta}, H_{CS}\rbrace,
    \label{multi}
\end{equation}
where $\chi^{\alpha}=(\chi^{1},\chi^{2})=(Q^{11},Q^{12})$ are the second class constraints. There are two second class constraints, then there are  two  Lagrange multipliers to be found, say  $(u_1, u_2)$. Hence, by using  (\ref{multi}), the following  multipliers arise 
\begin{eqnarray}
    u_{1}&=& C_{21}^{-1}\lbrace \chi^{2},H_{CS}\rbrace \\ \nonumber &=&2\pi^{12}+\partial_{2}\partial_{2}h_{00}-\partial_{1}\partial_{1}h_{00}+\frac{1}{2}\nabla^{2}h^{1}{}_{1}-\frac{1}{2}\nabla^{2}h^{2}{}_{2} +\frac{1}{2}\partial_{l}\partial_{2}h^{l}{}_{2}\nonumber \\ && -\frac{1}{2}\partial_{l}\partial_{1}h^{l}{}_{1} -2\partial_{2}G_{02}+2\partial_{1}G_{01}+v_{22}-v_{11},\\
    u_{2}&=& = C_{12}^{-1} \lbrace \chi^{1},H_{CS}\rbrace \\ \nonumber &=& -2\pi^{11}-2\left( \partial_{2}\partial_{1}h_{00}-\frac{1}{2}\nabla^{2}h_{12}+\frac{1}{2}\partial_{1}\partial_{l}h^{l}{}_{2}\right) +2\partial_{1}G_{02}+2\partial_{2}G_{01} -2v_{21},
\end{eqnarray}
and the extended Hamiltonian  takes the form 
\begin{equation}
    H_{ECS}= H_{CS}+u_{\alpha}\chi^{\alpha}=H_{CS}+u_{1}Q^{11}+u_{2}Q^{12}. 
    \label{eq:mult1}
\end{equation}
The extended Hamiltonian given in (\ref{eq:mult1}) is a  first class function. In fact,  its  algebra with all  first class constraints is given by 
\begin{eqnarray}
     \left\lbrace Q^{00},H_{ECS} \right\rbrace_{D}  &=&-\pi^{00}=-S^{0}, \\ 
     \left\lbrace Q^{0i},H_{ECS} \right\rbrace_{D} &=& -\delta^{i}{}_{2}\partial_{2}Q^{11}+\delta^{i}{}_{1}\partial_{1}Q^{11}+\delta^{i}{}_{2}\partial_{1}Q^{12}+\delta^{i}{}_{1}\partial_{2}Q^{12}\\ \nonumber && - \pi^{0i} - \frac{1}{4} \epsilon^{ij} \nabla^{2} h_{0j} - \frac{1}{2}\epsilon^{kj}\partial^{i}\partial_{k}h_{0j}= S^{i}, \\
     \left\lbrace S^{0},H_{ECS} \right\rbrace_{D}  &=& \partial_{1}\partial_{1}Q^{11}-\partial_{2}\partial_{2}Q^{11}+2\partial_{1}\partial_{2}Q^{12}-\epsilon^{ij}\partial_{j}\partial^{k}G_{ki}-\epsilon^{ij}\nabla^{2}\partial_{i}h_{0j}=\tilde{S}^{0},\\
     \left\lbrace U,H_{ECS} \right\rbrace_{D}  &=& -\tilde{V},\\
     \left\lbrace \tilde{V}^{i},H_{ECS} \right\rbrace_{D}  &=& 0,\\
     \left\lbrace \tilde{V},H_{ECS} \right\rbrace_{D}  &=& \partial_{1}\partial_{1}Q^{11}-\partial_{2}\partial_{2}Q^{11}+2\partial_{1}\partial_{2}Q^{12}-\epsilon^{ij}\partial_{j}\partial^{k}G_{ki}-\epsilon^{ij}\nabla^{2}\partial_{i}h_{0j}=\tilde{S}^{0},\\
     \left\lbrace \tilde{S}^{0},H_{ECS} \right\rbrace_{D}  &=& 0, \\
     \left\lbrace S^{i},H_{ECS} \right\rbrace_{D}  &=& \tilde{V}^{i},
\end{eqnarray}
where we observe that the algebra is closed. In this manner, we have performed a complete GLT analysis for the Chern-Simons term in the weak field context, and we have shown that the constraints are obtained with a correct structure  from the null vectors. \\
We finish this section with the calculation of the gauge transformations.  For this aim, we use the first class constraints and we define the generator of gauge transformations as \cite{11}
\begin{eqnarray}
\nonumber 
G&=& \int \Big\{ \varepsilon_{00}(y)Q^{00}(y)+ \varepsilon_{0i}(y)Q^{0i}(y)+ \varepsilon_{0}(y)S^{0}(y)+\varepsilon_{i}(y)S^{i}(y)+ \tilde{\varepsilon}_i(y)\tilde{V}^{i}(y)+\tilde{\varepsilon}(y)\tilde{V}(y) \\
&+& \psi (y)U(y)+ \tilde{\psi}_0(y) \tilde{S}^0(y) \Big\} d^2y,
\label{gen}
\end{eqnarray}
where $(\varepsilon_{00},\varepsilon_{0i}, \varepsilon_{0}, \varepsilon_{i}, \tilde{\varepsilon}_i, \tilde{\varepsilon},  \psi, \tilde{\psi}_0 )$ are gauge parameters. Hence, the gauge transformations of the canonical variables $h_{\mu \nu}$ are given by 
\begin{eqnarray}
\delta{h_{00}}(x)&=&\{ h_{00}(x), G\}_D=\varepsilon_0(x), \nonumber \\
\delta{h_{0i}}(x)&=& \{ h_{0i}(x), G\}_D=-\varepsilon_i(x), \nonumber \\
\delta{h_{ij}}(x)&=& \{ h_{ij}(x), G\}_D= -\frac{1}{2}(\partial_i \tilde{\varepsilon}_j(x)+ \partial_j \tilde{\varepsilon}_i(x))+ \delta_{ij}\tilde{\varepsilon}(x),
\end{eqnarray}
where (\ref{dbra}) have been used. If we fixing  the gauge parameters as $\varepsilon_0=2\partial_0\zeta_0, -\varepsilon_i=\partial_0\zeta_i+\partial_i\zeta_0,  \delta_{ij}\tilde{\varepsilon}= \frac{1}{2}(\partial_i \zeta_j+ \partial_j \zeta_i), -\tilde{\varepsilon}_i=\zeta_i$, then the gauge transformations are given by 
\begin{equation}
\delta h_{\mu \nu}= \partial_\mu \zeta_\nu+ \partial_\nu \zeta_\mu,  
\end{equation}
it is easy to see that  the action (\ref{eq:lcs}) and the equations of motion (see appendix B eq. (\ref{cot})) are invariant under these gauge transformations. 

\section{The GLT analysis for topologically massive gravity}
We have observed in previous sections,  that the GLT analysis for  the  higher-order  derivative Chern-Simons theory allowed us to know the complete structure of the constraints. Hence, in this section, we shall perform the GLT analysis for TMG theory and we will report the complete structure of the constraints; our study will complete  those results reported in \cite{15a}. \\
Now,  the action under study is given by 
\begin{equation}
    S[{g_{\mu \nu}}]=\int \Big \{ R\sqrt {-g} + \left[\frac{1}{\mu}\epsilon^{\lambda \mu \nu}\Gamma_{\lambda \sigma}^{\rho} (\partial_\mu \Gamma_{\rho \nu}^{\sigma} + \frac{2}{3} \Gamma_{\mu \xi}^{\sigma}\Gamma_{\nu \rho}^{\xi}) \right]\Big\} dx^3,
    \label{tmg}
\end{equation}
where $R$ is the Ricci tensor, $g$ is the determinant of the metric tensor and $\mu$  is a coupling constant \cite{ 11a}. If we consider the  week gravitational field  formalism,   the action (\ref{tmg}) is reduced  to   the well known Lagrangian for TMG which it is  given by  \cite{15a}
\begin{equation}
\begin{aligned}
    L_{TMG} = {} &  \frac{1}{4}\partial_{\lambda}h_{\mu \nu}\partial^{\lambda}h^{\mu \nu} - \frac{1}{4}\partial_{\lambda}h^{\mu}{}_{\mu}\partial^{\lambda}h^{\nu}{}_{\nu} + \frac{1}{2}\partial_{\lambda}h^{\lambda}{}_{\mu}\partial^{\mu}h^{\nu}{}_{\nu} - \frac{1}{2}\partial_{\lambda}h^{\lambda}{}_{\mu}\partial_{\nu}h^{\nu \mu}
    \\ & + \frac{1}{2 \mu}\epsilon^{\lambda \mu \nu} (\partial_\sigma h^{\rho}{}_{\lambda} \partial_\rho \partial_\mu h^{\sigma}{}_{\nu}-\partial_\sigma h^{\rho}{}_{\lambda} \partial^\sigma \partial_\mu h_{\rho \nu}).
\end{aligned}    
\label{eq:LTMG}    
\end{equation}
We have commented in previous sections  that  the action (\ref{eq:LTMG}) describes the propagation of a massive graviton on a Minkowski background \cite{11a}, and the analysis of this action always is   an interesting subject for studying due to the closeness with real gravity. \\
In this manner, by performing the 2+1 decomposition  we obtain  
\begin{equation}
\begin{aligned}
     L_{TMG} = {} & \frac{1}{4}\dot h_{ij} \dot h^{ij} - \frac{1}{4} \dot h^{i}{}_{i} \dot h^{k}{}_{k} + \dot h_{ij} \partial^{i}h^{0j} - \dot h^{i}{}_{i} \partial^{k} h^{0}{}_{k} - \frac{1}{2}\partial_{i}h_{0j}\partial^{i}h^{0j} \\ & - \frac{1}{4}\partial_{i}h_{jk}\partial^{i}h^{jk} +  \frac{1}{2}\partial_{i}h^{0}{}_{0}\partial^{i}h^{j}{}_{j} + \frac{1}{4}\partial_{i}h^{j}{}_{j}\partial^{i}h^{k}{}_{k} - \frac{1}{2}\partial_{i}h^{ij}\partial_{j}h^{0}{}_{0} \\ & - \frac{1}{2}\partial_{i}h^{ij}\partial_{j}h^{k}{}_{k} + \frac{1}{2}\partial_{i}h_{j0}\partial^{j}h^{0i} + \frac{1}{2}\partial_{i}h_{jk}\partial^{j}h^{ik} \\ & - \frac{1}{\mu}\epsilon^{ij}(\partial^k h_{0j}\Ddot{h}_{ki} + \partial_j h^{k}{}_{0}\Ddot{h}_{ki} - \frac{1}{2} \dot{h}^{k}{}_{j}\Ddot{h}_{ki} + \partial_j \partial^k h_{00} \dot{h}_{ki}+                \partial^k \partial_i h_{0j} \dot{h}_{k0} \\ & + \frac{1}{2} \nabla^2 h_{0j} \dot{h}_{0i} - \frac{1}{2} \nabla^2 h^{k}{}_{j} \dot{h}_{ki} + \frac{1}{2} \partial_k \partial_l h^{l}{}_{j} \dot{h}^{k}{}_{i} + \nabla^2 h_{00} \partial_i h_{0j} \\ & - \nabla^2 h^{k}{}_{0} \partial_i h_{kj} - \partial^l \partial_i h^{k}{}{j} \partial_k h_{l0}).
\end{aligned}
\label{eq:LLTMG}
\end{equation}
By following the GLT formalism, we introduce the variables  (\ref{eq:var}) into  (\ref{eq:LLTMG}) and we obtain 
\begin{equation}
\begin{aligned}
    L_{TMG}^{v}= {} & \frac{1}{4}G_{ij} G^{ij} - \frac{1}{4} G^{i}{}_{i} G^{k}{}_{k} + G_{ij} \partial^{i}h^{0j} - G^{i}{}_{i} \partial^{k} h^{0}{}_{k} - \frac{1}{2}\partial_{i}h_{0j}\partial^{i}h^{0j} \\ & - \frac{1}{4}\partial_{i}h_{jk}\partial^{i}h^{jk} +  \frac{1}{2}\partial_{i}h^{0}{}_{0}\partial^{i}h^{j}{}_{j} + \frac{1}{4}\partial_{i}h^{j}{}_{j}\partial^{i}h^{k}{}_{k} - \frac{1}{2}\partial_{i}h^{ij}\partial_{j}h^{0}{}_{0} \\ & - \frac{1}{2}\partial_{i}h^{ij}\partial_{j}h^{k}{}_{k} + \frac{1}{2}\partial_{i}h_{j0}\partial^{j}h^{0i} + \frac{1}{2}\partial_{i}h_{jk}\partial^{j}h^{ik} \\ & - \frac{1}{\mu}\epsilon^{ij}(\partial^k h_{0j} v_{ki} + \partial_j h^{k}{}_{0} v_{ki} - \frac{1}{2} G^{k}{}_{j} v_{ki} + \partial_j \partial^k h_{00} G_{ki} + \partial^k \partial_i h_{0j} G_{k0} \\ & + \frac{1}{2} \nabla^2 h_{0j} G_{0i} - \frac{1}{2} \nabla^2 h^{k}{}_{j} G_{ki} + \frac{1}{2} \partial^k \partial_l h^{l}{}_{j} G_{ki} + \nabla^2 h_{00} \partial_i h_{0j} \\ & - \nabla^2 h^{k}{}_{0} \partial_i h_{kj} - \partial^l \partial_i h^{k}{}_{j} \partial_k h_{l0}).
\end{aligned}
\label{eq:GvTGM}    
\end{equation}
hence, the canonical Hamiltonian for the theory under study is given by 
\begin{equation}
    \begin{aligned}
    H_{TMG}&={} \pi^{\mu \nu } \dot g_{\mu \nu } + P^{\mu \nu } \dot G_{\mu \nu } - L^v_{TMG} = \pi^{\mu \nu } G_{\mu \nu } + P^{\mu \nu } v_{\mu \nu } - L^v_{TMG} \\
     = {} & \pi^{00}G_{00} + P^{00}v_{00} +2\pi^{0k}G_{0k} + 2P^{0k}v_{0k} + \pi^{ki}G_{ki} + P^{ki}v_{ki} \\ & -\frac{1}{4}G_{ij} G^{ij} + \frac{1}{4} G^{i}{}_{i} G^{k}{}_{k} - G_{ij} \partial^{i}h^{0j} + G^{i}{}_{i} \partial^{k} h^{0}{}_{k} + \frac{1}{2}\partial_{i}h_{0j}\partial^{i}h^{0j} \\ & + \frac{1}{4}\partial_{i}h_{jk}\partial^{i}h^{jk} -  \frac{1}{2}\partial_{i}h^{0}{}_{0}\partial^{i}h^{j}{}_{j} - \frac{1}{4}\partial_{i}h^{j}{}_{j}\partial^{i}h^{k}{}_{k} + \frac{1}{2}\partial_{i}h^{ij}\partial_{j}h^{0}{}_{0} \\ & + \frac{1}{2}\partial_{i}h^{ij}\partial_{j}h^{k}{}_{k} - \frac{1}{2}\partial_{i}h_{j0}\partial^{j}h^{0i} - \frac{1}{2}\partial_{i}h_{jk}\partial^{j}h^{ik} \\ & + \frac{1}{\mu}\epsilon^{ij}(\partial^k h_{0j} v_{ki} + \partial_j h^{k}{}_{0} v_{ki} - \frac{1}{2} G^{k}{}_{j} v_{ki} + \partial_j \partial^k h_{00} G_{ki} + \partial^k \partial_i h_{0j} G_{k0} \\ & + \frac{1}{2} \nabla^2 h_{0j} G_{0i} - \frac{1}{2} \nabla^2 h^{k}{}_{j} G_{ki} + \frac{1}{2} \partial^k \partial_l h^{l}{}_{j} G_{ki} + \nabla^2 h_{00} \partial_i h_{0j} \\ & - \nabla^2 h^{k}{}_{0} \partial_i h_{kj} - \partial^l \partial_i h^{k}{}_{j} \partial_k h_{l0}).
    \end{aligned}
\label{eq:HTMG}
\end{equation}
and the fundamental Poisson brackets between the canonical variables will be 
\begin{eqnarray}
\nonumber 
    \left\lbrace h_{\alpha \beta},\pi^{\mu \nu}\right\rbrace &=&\frac{1}{2}\left(\delta^{\mu }{}_{\alpha} \delta^{\nu }{}_{\beta} + \delta^{\mu }{}_{\beta} \delta^{\nu }{}_{\alpha}\right), \\ 
    \left\lbrace G_{\alpha \beta},P^{\mu \nu}\right\rbrace&=&\frac{1}{2}\left(\delta^{\mu }{}_{\alpha} \delta^{\nu }{}_{\beta} + \delta^{\mu }{}_{\beta} \delta^{\nu }{}_{\alpha}\right), 
\label{eqbra2}
\end{eqnarray}
thus, from (\ref{eq:GvTGM}) we can identify the following primary constraints \cite{11b, 11}
\begin{equation}
    Q^{00}: \frac {\partial L_{TMG}^{v}}{\partial v_{00}} - P^{00}=0 \Rightarrow P^{00}\approx 0,
\label{eq:PC1}
\end{equation}
\begin{equation}
    Q^{0i}: \frac {\partial L_{TMG}^{v}}{\partial v_{0i}} - P^{0i}=0 \Rightarrow P^{0i} \approx 0, 
\label{eq:PC2}
\end{equation}
\begin{equation}
\begin{aligned}
    Q^{ki} {} & : \frac {\partial L_{TMG}^{v}}{\partial v_{ki}} - P^{ki}\\ & = - \left \{P^{ki} +\frac{1}{2\mu{}{}} \left [ \epsilon^{ij}(\partial^k h_{0j} + \partial_j h^{k}{}_{0} - \frac{1}{2} G^{k}{}_{j} ) + \epsilon^{kj}(\partial^i h_{0j} + \partial_j h^{i}{}_{0} - \frac{1}{2} G^{i}{}_{j} ) \right ] \right \} \approx 0.
\end{aligned}
\label{eq:PC3}
\end{equation}
Furthermore, the primary Hamiltonian is given in  usual  way 
\begin{equation}
       H^{1}_{TMG} := H_{TMG} + \Delta_{\mu \nu}Q^{{\mu \nu}},
\label{eq:HTMG1}
\end{equation}
where  $\Delta_{\mu \nu}$ are  identified as Lagrange multipliers. Now, from consistency of the primary constraints say 
\begin{equation}
    \dot Q^{\alpha \beta}=\left\lbrace Q^{\alpha \beta},H_{TMG}^{1} \right\rbrace=\left\lbrace Q^{\alpha \beta},H_{TMG} \right\rbrace + \Delta_{\mu \nu}\left\lbrace Q^{\alpha \beta},Q^{\mu \nu} \right\rbrace \approx 0.
\label{eq:PCEV}
\end{equation}
we identify the following secondary constraints 
\begin{equation}
S^{0}:   \pi^{00}  \approx 0 ,
\label{eq:SC1}
\end{equation}
\begin{equation}
 S^{i}:  \left ( \pi^{0i} + \frac{1}{4\mu}\epsilon^{ij}\nabla^2 h_{0j} + \frac{1}{2\mu}\epsilon^{lj} \partial^i \partial_l h_{0j} \right ) \approx 0,
\label{eq:SC2}
\end{equation}
and the following relations involving Lagrange multipliers
\begin{equation}
    \begin{split}
      S^{ik }& : \pi^{ik} - \frac{1}{2}G^{ik} + \frac{1}{2}\eta^{ik}G^{j}{}_{j} - \frac{1}{2} \left ( \partial^{i}h^{0k}+ \partial^{k}h^{0i} \right ) + \eta^{ik}\partial^{j}h^{0}{}_{j} \\ & + \frac{1}{2\mu} \bigg [ \epsilon^{ij} \left (\partial_j \partial^k h_{00} - \frac{1}{2} \nabla^2 h^{k}{}_{j} + \frac{1}{2} \partial^k \partial_r h^{r}{}_{j} \right ) + \epsilon^{kj} \left ( \partial_j \partial^i h_{00} - \frac{1}{2} \nabla^2 h^{i}{}_{j} + \frac{1}{2} \partial^i \partial_r h^{r}{}_{j} \right ) \bigg ] \\ &\quad - \frac{1}{2\mu} \left ( \epsilon^{lk}v^{i}_{l} + \epsilon^{li}v^{k}_{l} \right) - \frac{1}{2\mu} \bigg [ \epsilon^{ij} \left ( \partial^{k} G_{0j}+\partial_{j} G^{k}_{0} \right ) + \epsilon^{kj} \left ( \partial^{i} G_{0j}+\partial_{j} G^{i}_{0} \right ) \bigg ] \\ &\quad - \frac{1}{2\mu} \left (  \epsilon^{mk} \Delta^{i}_{m} + \epsilon^{mi} \Delta^{k}_{m}  \right ) \approx0.
\end{split}
\label{eq:LMEXP}
\end{equation}
From consistency of  $S^{0}$  we obtain a tertiary  constraint  
\begin{equation}
\dot S^{0} \rightarrow    \tilde{S}^0 : \frac{1}{2}\nabla^{2}h^{i}{}_{i} - \frac{1}{2}\partial_{j}\partial_{i}h^{ij} - \frac{1}{\mu}\left [\epsilon^{ij} \partial_{j} \partial^{k} G_{ki} + \epsilon^{ij} \nabla^{2} ( \partial_{i} h_{0j})  \right ]\approx0,
\end{equation}
and  preservation in time of $ S^{i}$   gives more relations involving Lagrange multipliers  
\begin{equation}
\begin{aligned}
\dot S^{i} \rightarrow    \tilde{S}^i :&  \frac{1}{2}\partial_{k}G^{ki} - \frac{1}{2}\partial^{i}G^{j}{}_{j} + \frac{1}{2}\nabla^{2}h^{0i} - \frac{1}{2}\partial_{j}\partial^{i}h^{0j} \\ & + \frac{\epsilon^{ik}}{2\mu} \left (\nabla^{2} G_{0k} + \partial^{l} \partial_{k} G_{0l} - \partial^{l}v_{lk} - \partial_{k}\nabla^{2}h_{00} \right ) \\ & + \frac{\epsilon^{kj}}{2\mu} \left ( \partial_{j} v^{i}{}_{k} + \nabla^{2}\partial_{k}h^{i}{}_{j} - \partial_{l} \partial^{i}\partial_{k}h^{l}{}_{j} + \partial^{i}\partial_{k}G_{0j} \right ) +\frac{1}{2\mu} \left( \epsilon^{ki}\partial^{l}\Delta_{lk} + \epsilon^{kj}\partial_{j}\Delta^{i}{}_{k}  \right) \approx0.
\end{aligned}
\label{eq:LMEXP2}
\end{equation}
Furthermore, from  consistency of $ \tilde{S}^0$ we do not obtain further constraints  but more relations between Lagrange multipliers
\begin{equation}
    \dot {\tilde{S}}^{0} = \frac{1}{2}\nabla^{2}G^{i}{}_{i} - \frac{1}{2}\partial^{j}\partial^{i}G_{ij} -\frac{1}{\mu}\epsilon^{ij} \left ( \partial_{j}\partial^{k}v_{ki} + \nabla^{2}\partial_{i}G_{0j} + \partial_{j}\partial^{k}\Delta_{ki} \right) \approx 0,
\end{equation}
Nonetheless, we can eliminate the Lagrange multipliers from (\ref{eq:LMEXP}) and (\ref{eq:LMEXP2}), thus,  a new constraint is obtained
\begin{equation}
\begin{split}
    \tilde{V}^{i} & = \partial_{k}\pi^{ik} + \frac{\epsilon^{kl}}{4\mu}\nabla^2\partial_kh^i{_{l}}- \frac{\epsilon^{kj}}{4\mu}\partial_k \partial^i\partial_lh^l{_{j}} \approx 0, 
\end{split}
\end{equation}
 the trace of  (\ref{eq:LMEXP}) remove the Lagrange multipliers and other new  constraint arise 
\begin{equation}
    \tilde{V}=\pi^{i}{}_{i} + \frac{1}{2}G^{j}{}_{j} + \partial^{i}h^{0}{}_{i} + \frac{1}{2\mu}\epsilon^{ij}\partial_{i}\partial_{l}h^{l}{}_{j} \approx 0.
\end{equation}
Moreover, from   the trace of  (\ref{eq:PC3}) emerges other constraint and   preservation in time  ends the search of further constraints 
\begin{eqnarray}
        & Q^{ik}\eta_{ik}& = P^{i}{}_{i}\approx 0, \nonumber \\
        &\dot P^{i}{}_{i}& = \left\lbrace P^{i}{}_{i},H^{1}_{CS} \right\rbrace = \tilde{V} \approx 0.
\end{eqnarray}
Therefore the complete set of constraints is given by 
\begin{eqnarray}
&Q^{00}&:P^{00} \approx0, \nonumber \\
&Q^{0i}&:P^{0i} \approx0, \nonumber \\
&Q^{ik}&:P^{ki} +\frac{\epsilon^{ij}}{2\mu}  \partial^k h_{0j} + \frac{ \epsilon^{ij}}{2\mu}\partial_j h^{k}{}_{0} - \frac{\epsilon^{ij}}{4\mu} G^{k}{}_{j}  + \frac{\epsilon^{kj}}{2\mu} \partial^i h_{0j} + \frac{\epsilon^{kj}}{2\mu}\partial_j h^{i}{}_{0} - \frac{\epsilon^{kj}}{4\mu} G^{i}{}_{j} \approx0,  \nonumber \\
&S^{0}& :   \pi^{00}  \approx 0, \nonumber \\
&S^{i}&:    \pi^{0i} + \frac{\epsilon^{ij}}{4\mu}\nabla^2 h_{0j} + \frac{\epsilon^{lj}}{2\mu}\partial^i \partial_l h_{0j}  \approx 0, \nonumber\\
 &\tilde{S}^0& : \frac{1}{2}\nabla^{2}h^{j}{}_{j} - \frac{1}{2}\partial_{i}\partial_{j}h^{ij} - \frac{1}{\mu}\epsilon^{ij} \partial_{j} \partial^{k} G_{ki} -\frac{1}{\mu} \epsilon^{ij} \nabla^{2}  \partial_{i} h_{0j} \approx 0, \nonumber \\
  &\tilde{V}^{i} & :\partial_{k}\pi^{ik} + \frac{\epsilon^{kl}}{4\mu}\nabla^2\partial_kh^i{_{l}}- \frac{\epsilon^{kj}}{4\mu}\partial_k \partial^i\partial_lh^l{_{j}} \approx 0, \nonumber \\
 & \tilde{V}&:\pi^i{_{i}} +\frac{1}{2}G^{j}{}_{j} + \partial_{i}h^{0i} + \frac{\epsilon^{ij}}{2\mu} \partial_i \partial_lh^l{_{j}} \approx 0, \nonumber \\
 &U&:P^{i}{}_{i} \approx 0.
\end{eqnarray}

Now, we will separate  the constraints  into first class and second class. For this aim, we calculate the matrix, say   $W^{\alpha \beta}$,  whose entries are given by the Poisson brackets between all constraints, this is 
\begin{equation}
W = \bordermatrix{
                   & Q^{00}    & Q^{0l}   & Q^{lm}   & S^{ 0}   & S^{ l}   & \tilde{S}^{0}   & \tilde{V}^{l}   & \tilde{V}  &U    \cr
    Q^{00}         & 0         & 0        & 0        & 0        & 0        & 0               & 0               & 0          &0    \cr
    Q^{0i}         & 0         & 0        & 0        & 0        & 0        & 0               & 0               & 0          &0    \cr
    Q^{ik}         & 0         & 0        & \left\lbrace Q^{ik},Q^{lm} \right\rbrace        & 0        & \left\lbrace Q^{ik},S^{l} \right\rbrace & \left\lbrace Q^{ik},\tilde{S}^{0} \right\rbrace                & 0               & \left\lbrace Q^{ik},\tilde{V} \right\rbrace       &U     \cr
    S^{ 0}         & 0         & 0        & 0        & 0        & 0        & 0               & 0               & 0          &0    \cr
    S^{ i}         & 0         & 0        & \left\lbrace S^{i},Q^{lm} \right\rbrace        & 0        & \left\lbrace S^{i},S^{l} \right\rbrace        & \left\lbrace S^{i},\tilde{S}^{0} \right\rbrace               & 0               & \left\lbrace S^{i},\tilde{V} \right\rbrace     &0       \cr
    \tilde{S}^{0}  & 0         & 0        & \left\lbrace \tilde{S}^{0},Q^{lm} \right\rbrace        & 0        & \left\lbrace \tilde{S}^{0},{S}^{l} \right\rbrace         & 0               & 0               & \left\lbrace \tilde{S}^{0},\tilde{V} \right\rbrace      &0     \cr
    \tilde{V}^{i}  & 0         & 0        & 0        & 0        & 0        & 0               & 0            & 0         &0        \cr
    \tilde{V}      & 0         & 0        & \left\lbrace \tilde{V},Q^{lm} \right\rbrace         & 0        & \left\lbrace \tilde{V},S^{l} \right\rbrace         & \left\lbrace \tilde{V},\tilde{S}^{0} \right\rbrace                & 0            & 0        & \left\lbrace \tilde{V},U \right\rbrace        \cr
    U         & 0         & 0        & 0        & 0        & 0        & 0               & 0               &  \left\lbrace U, \tilde{V} \right\rbrace       &0        \cr
            }, 
            \label{eq:WTMG}
\end{equation}
where the nontrivial brackets are given by 
\begin{eqnarray}
& \left\lbrace Q^{ik},Q^{lm} \right\rbrace & = \frac{1}{4\mu} \left (  \epsilon^{mi}\eta^{lk} + \epsilon^{li}\eta^{mk} + \epsilon^{lk}\eta^{mi} + \epsilon^{mk}\eta^{li} \right )\delta^{2}(x-y), \nonumber \\
& \left\lbrace Q^{ik},S^{l} \right\rbrace & =  \frac{1}{4\mu} \left ( \epsilon^{il}\partial^{k} + \epsilon^{kl}\partial^{i} + \epsilon^{ij}\eta^{kl}\partial_{j} + \epsilon^{kj}\eta^{il}\partial_{j} \right )\delta^{2}(x-y), \nonumber \\
& \left\lbrace Q^{ik},\tilde{S}^{0} \right\rbrace & = \frac{1}{2\mu} \left ( \epsilon^{ij}\partial_{j}\partial^{k} + \epsilon^{kj}\partial_{j}\partial^{i} \right ) \delta^{2}(x-y), \nonumber \\
& \left\lbrace Q^{ik},\tilde{V} \right\rbrace & = -\frac{1}{2}\eta^{ik}\delta^{2}(x-y) , \nonumber \\
& \left\lbrace S^{i},S^{l} \right\rbrace & = \frac{1}{4\mu} \left ( \epsilon^{il}\nabla^{2} + \epsilon^{ij}\partial^{l}\partial_{j} + \epsilon^{jl}\partial^{i}\partial_{j} \right )\delta^{2}(x-y), \nonumber \\
& \left\lbrace S^{i},\tilde{V} \right\rbrace & = -\frac{1}{2}\partial^{i}\delta^{2}(x-y) , \nonumber \\
& \left\lbrace \tilde{S}^{0},\tilde{V} \right\rbrace & = \frac{1}{2}\nabla^{2}\delta^{2}(x-y) , \nonumber \\
& \left\lbrace S^{i},\tilde{S}^{0} \right\rbrace & = \frac{1}{2\mu}\epsilon^{ji}\nabla^{2}\partial_{j}\delta^{2}(x-y), \nonumber \\
& \left\lbrace \tilde{V},U \right\rbrace & = \delta^{2}(x-y),
\end{eqnarray}
hence, after a long algebraic manupulations,  we observe that tha matrix (\ref{eq:WTMG})  has a rank= 4 and nine null vectors. From the null vectors we identify the following first class constraints 
    \begin{eqnarray}
&Q^{00}&:P^{00} \approx0, \nonumber \\
&Q^{0i}&:P^{0i} \approx0, \nonumber \\
&S^{0}& :   \pi^{00}  \approx 0, \nonumber \\
&S^{i}&:  \partial_{k}P^{ki} - \pi^{0i} + \frac{\epsilon^{ij}}{4\mu}\nabla^{2}h_{0j} + \frac{\epsilon^{ij}}{2\mu}\partial_{j}\partial^{k}h_{k0} - \frac{\epsilon^{ij}}{4\mu}\partial^{k}G_{kj} - \frac{\epsilon^{kj}}{4\mu}\partial_{k}G^{i}_{j} \approx0 \nonumber\\
&\tilde{S}^0& \partial_{i}\partial_{k}P^{ik} + \frac{1}{2}\nabla^{2}h^{j}_{j}-\frac{1}{2}\partial_{i}\partial_{j}h^{ij} + \frac{\epsilon^{ij}}{2\mu}\partial_{i}\partial_{k}G^{k}{}_{j} \approx0 \nonumber\\
&\tilde{V}^{i} & :\partial_{k}\pi^{ik} + \frac{\epsilon^{kl}}{4\mu}\nabla^2\partial_kh^i{_{l}}- \frac{\epsilon^{kj}}{4\mu}\partial_k \partial^i\partial_lh^l{_{j}} \approx 0, 
\end{eqnarray}
and the rank  implies  the following four second class constraints 
\begin{eqnarray}
& Q^{11}& = P^{11} + \frac{1}{\mu}\partial^{1}h_{02} + \frac{1}{\mu}\partial_{2}h^{1}{}_{0} -\frac{1}{2\mu}G^{1}{}_{2} \approx 0, \nonumber\\
& Q^{12}& = P^{12} + \frac{1}{\mu}\partial^{2}h_{02} - \frac{1}{\mu}\partial^{1}h_{01} + \frac{1}{4\mu}G^{1}{}_{1} - \frac{1}{4\mu}G^{2}{}_{2}  \approx 0, \nonumber \\
 &\tilde{V}&:\pi^i{_{i}} +\frac{1}{2}G^{j}{}_{j} + \partial_{i}h^{0i} + \frac{\epsilon^{ij}}{2\mu} \partial_i \partial_lh^l{_{j}} \approx 0, \nonumber\\
 &U&: P^{i}{}_{i} \approx 0.
\end{eqnarray}
In this manner, the complete set of constraints allow us to carry out the counting of physical degrees of freedom as follows; there are 24 canonical variables $(h_{\mu\nu}, G_{\mu\nu}, \pi^{\mu\nu}, P^{\mu\nu})$,  four  second class constraints $(Q^{11}, Q^{12}, \tilde V, U)$ and nine  first class constraints $(Q^{00}, Q^{0i}, S^{0}, S^{i}, \tilde S^{0}, \tilde V^{i})$, thus 
\begin{equation}
DOF= \frac{1}{2}\left[ \left ( 24 \right ) -2\left ( 9\right ) - \left (4\right ) \right] = 1,
\end{equation}
as  expected \cite{10, 11a, 15a}. From the second class constraints we can construct the Dirac brackets between the canonical variables. The nontrivial Dirac's brackets are given by 
\begin{eqnarray}
&\left\lbrace h_{00},\pi^{00} \right\rbrace_{D}& = \delta^{2}(x-y), \nonumber \\
&\left\lbrace h_{0i},\pi^{0l} \right\rbrace_{D}& = \frac{1}{2}\delta_{l}{}^{i}\delta^{2}(x-y), \nonumber \\
&\left\lbrace h_{ij},\pi^{lm} \right\rbrace_{D}& = \frac{1}{2} \left( \delta_{i}{}^{l}\delta_{j}{}^{m} + \delta_{i}{}^{m}\delta_{j}{}^{l}\right) \delta^{2}(x-y), \nonumber \\
&\left\lbrace h_{ij},G_{lm} \right\rbrace_{D}& = -\eta_{ij}\eta_{lm}\delta^{2}(x-y), \nonumber \\
&\left\lbrace \pi^{0i},\pi^{0l} \right\rbrace_{D}& = -\frac{1}{2\mu} \epsilon^{il}\nabla^{2} \delta^{2}(x-y), \nonumber \\
&\left\lbrace \pi^{0i},G_{lm} \right\rbrace_{D}& = \frac{1}{2}\left( \delta^{i}{}_{m}\partial_{l}+\delta^{i}{}_{l}\partial_{m} \right)\delta^{2}(x-y), \nonumber \\
&\left\lbrace \pi^{0i},P^{lm} \right\rbrace_{D}& = -\frac{1}{8\mu}\left(  \epsilon^{ln}\eta^{im} + \epsilon^{mn}\eta^{il} + \epsilon^{mi}\eta^{ln} + \epsilon^{li}\eta^{mn}  \right)\partial_{n} \delta^{2}(x-y), \nonumber \\
&\left\lbrace \pi^{ij},G_{lm} \right\rbrace_{D}& = \frac{1}{4\mu}\eta_{lm}\left( \epsilon^{nj}\partial_{n}\partial^{i}+\epsilon^{ni}\partial_{n}\partial^{j} \right)\delta^{2}(x-y), \nonumber \\
&\left\lbrace G_{00},P^{00} \right\rbrace_{D}& = \delta^{2}(x-y), \nonumber \\
&\left\lbrace G_{0i},P^{0l} \right\rbrace_{D}& = \frac{1}{2}\delta_{i}{}^{l}\delta^{2}(x-y), \nonumber \\
&\left\lbrace G_{ij},G_{lm} \right\rbrace_{D}& =\frac{\mu}{4}\left(  \epsilon_{il}\eta_{jm}+\epsilon_{im}\eta_{jl}+\epsilon_{jl}\eta_{im}+\epsilon_{jm}\eta_{il}\right)\delta^{2}(x-y) ,\nonumber \\
&\left\lbrace G_{ij},P^{lm} \right\rbrace_{D}& = \frac{1}{4}\left( \delta^{l}{}_{i}\delta^{m}{}_{j}+\delta^{l}{}_{j}\delta^{m}{}_{i}-\eta^{lm}\eta_{ij} \right)\delta^{2}(x-y) \nonumber \\
&\left\lbrace P^{ij},P^{lm} \right\rbrace_{D}& = \frac{1}{16\mu}\left( \epsilon^{il}\eta^{mj} + \epsilon^{im}\eta^{jl} + \epsilon^{jl}\eta^{mi} + \epsilon^{jm}\eta^{il} \right)\delta^{2}(x-y),
\end{eqnarray}
where we can observe that these brackets coincide with those  reported in \cite{15a} where the Ostogradski approach was used. Furthermore, the extended action   is given by 
\begin{equation}
    S_{EXT}\left[ h_{\mu \nu}, \pi^{\mu \nu}, G_{\mu \nu}, P^{\mu \nu}, \lambda_{\alpha}, w_{\alpha} \right] = \int \left( \pi^{\mu \nu}\dot{h}_{\mu \nu} + P^{\mu \nu}\Ddot{h}_{\mu \nu}-H_{ETMG}-\lambda_{\alpha}\gamma^{\alpha}-w_{\alpha}\chi^{\alpha}  \right) \,d^{3}x, \nonumber
\end{equation}
where $H_{ETMG}$ is the extended  Hamiltonian, $\gamma^{\alpha}$ are the first class constraints, $\chi^{\alpha}$ are the second  class constraints; $\lambda_{\alpha}$ and $w_{\alpha}$ are Lagrange  multipliers enforcing the first class and second class  constraints respectively. The extended Hamiltonian is given by 
\begin{equation}
    H_{ETMG}= H_{TMG}+u_{\alpha}\chi^{\alpha}=H_{TMG}+u_{1}Q^{11}+u_{2}Q^{12}+u_{3}\tilde{V}+u_{4}U, 
    \label{eq:mult}
\end{equation}
where the u's are four Lagrange multipliers that must be identified  because they are associated with the four second class constraints. Hence, the $ u$'s are given by 
\begin{equation}
    u_{\alpha}= C^{-1}_{\beta \alpha}\lbrace \chi^{\beta}, H_{TMG}\rbrace ,
\label{multi2}
\end{equation}
where $\chi^{\alpha}=(Q^{11},Q^{12},\tilde{V},U)$. In this manner, from (\ref{multi2}) we obtain the following expresions for the   multipliers 
\begin{eqnarray}
u_{1}&=& C_{21}^{-1}\lbrace \chi^{1},H_{TMG}\rbrace  \nonumber \\ &=& ( 2\mu\pi^{12} -\mu G^{12} -\mu\partial^{1}h^{02}-\mu\partial^{2}h^{01}+\partial^{2}\partial_{2}h_{00}-\partial^{1}\partial_{1}h_{00}+\frac{1}{2}\nabla^{2}h^{1}{}_{1}-\frac{1}{2}\nabla^{2}h^{2}{}_{2}\ \nonumber \\ && +\frac{1}{2}\partial_{l}\partial^{2}h^{l}{}_{2} -\frac{1}{2}\partial_{l}\partial^{1}h^{l}{}_{1} -2\partial^{2}G_{02}+2\partial^{1}G_{01}+v^{2}{}_{2}-v^{1}{}_{1})\delta^{2}(x-y), \\
u_{2}&=&C_{12}^{-1} \lbrace \chi^{1},H_{TMG}\rbrace + C_{42}^{-1}\lbrace \chi^{4},H_{TMG}\rbrace \nonumber \\ &=& (-2\mu\pi^{11} -\mu G^{22}-2\mu \partial^{2}h^{02}-2\partial_{2}\partial^{1}h_{00}+\nabla^{2}h^{1}{}_{2}-\partial^{1}\partial_{l}h^{l}{}_{2} \nonumber\\&& + 2\partial^{1}G_{02}+2\partial_{2}G^{1}{}_{0} -2v^{1}_{2} +\mu \tilde{V})\delta^{2}(x-y), \\
u_{3}&=& C_{43}^{-1} \lbrace \chi^{4},H_{TMG}\rbrace = -\tilde{V}\delta^{2}(x-y), \\
u_{4}&=&  C_{24}^{-1}\lbrace \chi^{2},H_{TMG}\rbrace +C_{34}^{-1}\lbrace \chi^{3},H_{TMG}\rbrace \nonumber\\ &=&(-\mu\pi^{12} +\frac{\mu}{2} G^{12} +\frac{\mu}{2}\partial^{1}h^{02}+\frac{\mu}{2}\partial^{2}h^{01}-v^{2}_{2}-\partial_{2}\partial^{2}h_{00}-\frac{1}{4}\nabla^{2}h^{1}{}_{1}+\frac{1}{4}\nabla^{2}h^{2}{}_{2} \nonumber \\ && -\frac{1}{4}\partial_{l}\partial^{2}h^{l}{}_{2} +\frac{1}{4}\partial_{l}\partial^{1}h^{l}{}_{1} +2\partial^{1}G_{01}-\frac{1}{\mu}\epsilon^{ij}\partial_{i}\partial^{k}G_{kj}+\frac{1}{\mu}\epsilon^{ij}\partial_{i}\nabla^{2}h_{j0} )\delta^{2}(x-y), 
\end{eqnarray}
the substitution of these multipliers in the extended Hamiltonian defines a first class function. In fact, the Dirac  algebra between the first class constraints and the extended Hamiltonian is given by 
\begin{eqnarray}
     \left\lbrace Q^{00},H_{ETMG} \right\rbrace_{D} &=&-S^{0}, \\ 
     \left\lbrace Q^{0i},H_{ETMG} \right\rbrace_{D} &=& S^{i}, \\
     \left\lbrace S^{0},H_{ETMG} \right\rbrace_{D}  &=& \tilde{S}^{0},\\
     \left\lbrace \tilde{V}^{i},H_{ETMG} \right\rbrace_{D}  &=&0 ,\\
     \left\lbrace \tilde{S}^{0},H_{ETMG} \right\rbrace_{D} &=& 0, \\
     \left\lbrace S^{i},H_{ETMG} \right\rbrace_{D}  &=& \tilde V^{i} ,
\end{eqnarray}
therefore, with this result, we have constructed  a first class extended Hamiltonian as expected. We have observed that the constraints  in this work are not the same at all to  those reported in \cite{15a}, however, the results of the Dirac brackets are equivalent. In this manner, we have presented a new alternative canonical analysis for TMG that extends those reported in the literature.

\section{Conclusions}
In this paper, a detailed GLT analysis of higher-order Chern-Simons and TMG theories in the perturbative context has been performed. With respect to higher-order Chern-Simons theory, we have reported all the complete structure of the constraints, as far as we know, a complete analysis of the constraints of this theory in the weak field context has not been reported. We found the extended Hamiltonian and we showed by means the Dirac algebra that it is of first-class. On the other hand, from the analysis of TMG we have reported a new structure of the constraints. We obtained the constraints by means of the null vectors and we showed that our analysis is consistent; we constructed a first-class extended Hamiltonian and the Dirac algebra between it and the first-class constraints is closed. In this manner, we observed that GLT formalism is an elegant and pragmatic scheme for analyzing higher-order theories.\\
It is worth emphasizing, that the GLT framework can be applied to other physical systems. In fact, it can be used  in others theories  where the Ostrogradski framework is not easy to develop. In this sense, the results of this paper are the basis for future works where the GLT framework will show advantages for the analysis of the constraints,  for instance,  in higher order modifications of gravity.  However, all  these ideas are in progress and will be reported soon  \cite{19}. 
\section{Appendix A}
In general, the higher order Lagrangian  (\ref{eq:lcs}) has the form 
\begin{equation}
L_{CS}=L_{CS}\Big(h_{\alpha\beta}, \partial_0h_{\alpha\beta}, \partial_ih_{\alpha\beta}, \partial_{0}\partial_0h_{\alpha\beta}, 2\partial_{0}\partial_ih_{\alpha\beta}, \partial_{i}\partial_jh_{\alpha\beta}  \Big), 
\label{A1}
\end{equation}
hence, the equations of motion obtained from (\ref{A1}) are given by 
\begin{eqnarray}
&&\frac{\partial L_{CS}}{\partial h_{\alpha \beta}} -\partial_0 \frac{\partial L_{CS}}{\partial (\partial_0 h_{\alpha \beta})}- \partial_i \frac{\partial L_{CS}}{\partial (\partial_i h_{\alpha \beta})} + \partial_0\partial_0 \frac{\partial L_{CS}}{\partial (\partial_0\partial_0 h_{\alpha \beta})} \nonumber \\ 
&+&  \partial_0\partial_i \frac{\partial L_{CS}}{\partial (\partial_i\partial_0 h_{\alpha \beta})} + \partial_i\partial_j \frac{\partial L_{CS}}{\partial (\partial_i\partial_j h_{\alpha \beta})}=0. 
\label{B1}
\end{eqnarray}
On the other hand, the introduction of the momenta  $(\pi ^{\mu \nu},  P^{\mu \nu})$   allows us to rewrite the Lagrangian  (\ref{A1}) as  (see Eq. (\ref{eq:actioncs}))
  \begin{equation}
  S= \int{\left[ L_{CS}^{v} + \pi ^{\mu \nu}(\dot h_{\mu \nu} - G_{\mu \nu}) + P^{\mu \nu}(\dot G_{\mu \nu} - v_{\mu \nu})\right]{d}^{3}x},
  \label{B2}
  \end{equation}
  where $L_{CS}^{v}= L_{CS}^{v}\Big(h_{\alpha\beta}, G_{\alpha\beta}, \partial_ih_{\alpha\beta}, v_{\alpha\beta}, \partial_{i}G_{\alpha\beta}, \partial_{i}\partial_jh_{\alpha\beta}  \Big)$, $G_{\alpha \beta}= \dot{h}_{\alpha \beta}$ and $\dot{G}_{\alpha \beta}=v_{\alpha \beta}$. Hence, the variation of (\ref{B2}) respect all dynamical variables is given by 
  \begin{eqnarray}
  \frac{\delta S}{\delta h_{\alpha \beta}} &=&  \frac{\partial L_{CS}^{v}}{\partial  h_{\alpha \beta}} - \partial_i \frac{\partial L_{CS}^{v}}{\partial (\partial_i h_{\alpha \beta})}+ \partial_i \partial_j \frac{\partial L_{CS}^{v}}{\partial (\partial_i \partial_j h_{\alpha \beta})}-\dot{\pi}^{\alpha \beta}=0,  \label{95}\\
  \frac{\delta S}{\delta G_{\alpha \beta}} &=&   \frac{\partial L_{CS}^{v}}{\partial  G_{\alpha \beta}}- \partial_i \frac{\partial L_{CS}^{v}}{\partial (\partial_i G_{\alpha \beta})}- \pi^{\alpha \beta} - \dot{P}^{\alpha \beta}=0, \label{96}\\
  \frac{\delta S}{\delta v_{\alpha \beta}} &=&  \frac{\partial L_{CS}^{v}}{\partial  v_{\alpha \beta}}- P^{\alpha \beta}=0,  \label{97} \\
  \frac{\delta S}{\delta \pi^{\alpha \beta}} &=& \dot{h}_{\alpha \beta} - G_{\alpha \beta}=0, \label{98}\\
    \frac{\delta S}{\delta P^{\alpha \beta}} &=& \dot{G}_{\alpha \beta}- v_{\alpha \beta}=0. \label{99}
  \end{eqnarray}
  The equations (\ref{95}-\ref{99}) are equivalent to (\ref{B1}). In fact, from time differentiation   of (\ref{97}) and taking  (\ref{96}) into account,   we obtain
  \begin{equation}
  \frac{\partial L_{CS}^{v}}{\partial  G_{\alpha \beta}}- \partial_i \frac{\partial L_{CS}^{v}}{\partial (\partial_i G_{\alpha \beta})}- \pi^{\alpha \beta}- \partial_0 \frac{\partial L_{CS}^{v}}{\partial v_{\alpha \beta}}=0,
  \label{100}
  \end{equation}
  now, substituting the  time derivative of (\ref{100})  into (\ref{95}) the following arise 
  \begin{eqnarray}
 && \frac{\partial L_{CS}^{v}}{\partial  h_{\alpha \beta}} - \partial_i \frac{\partial L_{CS}^{v}}{\partial (\partial_i h_{\alpha \beta})}+ \partial_i \partial_j \frac{\partial L_{CS}^{v}}{\partial (\partial_i \partial_j h_{\alpha \beta})}-  \partial_0 \frac{\partial L_{CS}^{v}}{\partial  G_{\alpha \beta}}+ \partial_0 \partial_i \frac{\partial L_{CS}^{v}}{\partial (\partial_i G_{\alpha \beta})}+ \partial_0\partial_0 \frac{\partial L_{CS}^{v}}{\partial v_{\alpha \beta}}=0,
 \label{101}
  \end{eqnarray} 
  finally, if the solutions of (\ref{98}-\ref{99})  are substituted into (\ref{101}), then (\ref{B1}) and (\ref{101}) are equivalents \cite{22, 23}. 
 \section{Appendix B}
In this appendix we will resume the difference  between the Chern-Simons and TMG theories. In fact, the equations of motion obtained from  the Chern-Simons action (\ref{eq:lcs}) are given by 
\begin{equation}
C^{(L)}_{\mu \nu }=\epsilon_{\mu}{^{\alpha \beta}\partial_\alpha R^{(L)}_{\beta \nu} } + \epsilon_{\nu}{^{\alpha \beta}} \partial_\alpha R^{(L)}_{\beta \mu}=0, 
\label{cot}
\end{equation}
where $C^{(L)}_{\mu \nu }$ is the linear Cotton tensor and $R^{(L)}_{\beta \nu}$ is the linear Ricci tensor given by 
\begin{equation}
R^{(L)}_{\mu \nu}=\frac{1}{2} \big( -\Box h_{\mu \nu} - \partial_\mu \partial_\nu h+ \partial^\sigma \partial_\nu h_{\sigma \mu}  + \partial^\sigma \partial_\mu h_{\sigma \nu} \big).  
\end{equation}
It is well-known that Cotton tensor vanishes  if and only if
the 3-dimensional metric tensor is conformally flat, and this shows  a  difference between Chern-Simons theory and TMG. In fact,  Chern-Symons theory does not describe propagation of degrees of freedom, there is not gravity. \\
On the other hand,  the equations of motion  obtained from  (\ref{eq:LTMG}) are given by 
\begin{equation}
G_{\alpha \nu}^{(L)} + \frac{1}{\mu} C^{(L)}_{\alpha \nu }=0,
\label{lin}
\end{equation}
where $G_{\alpha \nu}^{(L)} = R^{(L)}_{\alpha \nu} - \frac{1}{2}g_{\alpha \nu} R^{(L)}$  is the linear Einstein tensor.  By using the Coulomb gauge $\partial^\alpha h_{\alpha \beta}=0$ and the traceless condition $h{^{\alpha}}_\alpha=0$,  the equation (\ref{lin}) can be reduced to  the branch of the massive sector given by 
\begin{equation}
h_{\alpha \nu}+ \frac{1}{\mu} \epsilon_{\alpha}{^{\rho \beta}}h_{\beta \nu}=0, 
\label{lin2}
\end{equation}
if we contract the equation  (\ref{lin2}) with $(\eta{^\nu}_{\rho} - \frac{1}{\mu} \epsilon_{\rho}{^{ \alpha \nu}}\partial_\alpha)$ we obtain 
\begin{equation}
\big(\Box - \mu^2 \big) h_{\mu \nu}=0.
\label{mass2}
\end{equation}
Equation (96) describes the propagation of one degree of freedom a massive spin-2 graviton with mass  $m=\sqrt{\mu^2}$  \cite{20}, now the solutions are even  more general  than Einstein's metrics \cite{21}. In this respect,  the TMG theory is  attractive because it is close  to real gravity. 

\end{document}